\documentclass[aps,prb,showpacs,twocolumn,superscriptaddress,floats,floatfix]{revtex4}
\usepackage{amssymb}
\usepackage{amsbsy}
\usepackage{amsmath}
\usepackage{epsfig}

\begin{document}
\title {Superfluid-insulator transitions of two-species Bosons in an optical lattice}

\author{A. Isacsson}
\affiliation{NORDITA, Blegdamsvej 17, Copenhagen \O, DK-2100, Denmark}
\affiliation{Department of Physics, Yale University, P.O. Box 208120,
             New Haven, CT 06520-8120}
\author{Min-Chul Cha}
\affiliation{Department of Applied
Physics, Hanyang University, Ansan, Kyunggi-do 426-791, Korea}
\affiliation{Department of Physics, Yale University, P.O. Box
208120, New Haven, CT 06520-8120} 

\author{K. Sengupta}
\affiliation{Harish-Chandra Research Institute, Chhatnag Road, Jhunsi,
Allahabad 211019, India. }
\affiliation{Department of Physics, Yale University, P.O. Box
208120, New Haven, CT 06520-8120} 

\author{S.M. Girvin}
\affiliation{Department of Physics, Yale University, P.O. Box 208120,
New Haven, CT 06520-8120}

\date{\today}

\begin{abstract}

We consider a realization of the two-species bosonic Hubbard model
with variable interspecies interaction and hopping strength. We
analyze the superfluid-insulator (SI) transition for the relevant
parameter regimes and compute
the ground state phase diagram for odd filling at commensurate
densities. We find that in contrast to the even commensurate
filling case, the superfluid-insulator transition occurs with (a)
simultaneous onset
of superfluidity of both species or (b) coexistence of
Mott insulating state of one species and superfluidity of the
other or, in the case of unit filling, (c) complete depopulation
of one species.  The superfluid-insulator transition can be first
order in a large region of the phase diagram. We develop a
variational mean-field method which takes into account the effect
of second order quantum fluctuations on the superfluid-insulator
transition and corroborate the mean-field phase diagram using a
quantum Monte Carlo study.

\end{abstract}

\maketitle
\section{Introduction}

Experiments with ultracold atoms have achieved reversible tuning
of bosonic atoms between superfluid (SF) and Mott insulating (MI) states by
varying the strength of periodic potential produced by standing
laser light \cite{Bloch1,Kasevich}. The physics of such ultracold
atoms in the Mott insulating state can be described by a bosonic
Hubbard model, well known in context of other condensed matter
systems \cite{Sachdev1}. However, ultracold atoms in optical
lattices offer much better control over microscopic parameters of
the model. Consequently, it is possible to explore parameter
regimes which are not available in other analogous condensed
matter systems.

Recently, experiments involving internal states of these atoms have
been carried out \cite{Bloch2, Legget}. In particular, in Ref.\
\onlinecite{Bloch2}, the two hyperfine states
($\left|F=2,m_F=-2\right> \equiv \left|1\right>$ and
$\left|F=1,m_F=-1\right> \equiv \left|2\right>$) of $\rm ^{87}Rb$
atoms have been used to create entangled states between atoms in
different wells of the optical lattice. In these experiments, a
$\pi/2$ pulse is applied to bosons originally in one of the two
hyperfine states (say $\left|1\right>$), leaving them in eigenstates
of $\sigma_x$ ($[\left|1\right>+\left|2\right>]/\sqrt{2}$), where
the $\sigma$ denote Pauli matrices corresponding to the two
hyperfine states.

To envisage how such experimental systems are relevant for
realization of a two species Bose-Hubbard model, consider an optical
lattice created using elliptically polarized light with polarization
angle $\theta$. Since the spin states with $m_s = \pm 1/2$ see
potentials $ V_{\pm} = V_0 \sin^2(kx \pm \theta)$, the hyperfine
states $\left|1\right>$ and $\left|2\right>$ experience potentials
$V_{1(2)}$ given by (see Refs.\ \onlinecite{Bloch2,Zoller,Brennen}
for details)
\begin{eqnarray}
V_{1} &=& V_0 \sin^2(kx+\theta) \nonumber \\
V_2 &=& \frac{V_0}{4} \left( \sin^2(kx+\theta) +
3\sin^2(kx-\theta)\right).
\end{eqnarray}
Consequently, a change in the polarization angle $\theta$ is
equivalent to a relative shift of the lattices with respect to each
other. Since the interaction between the bosons is short-ranged,
such a shift can be used to control the inter-species interaction
$U'$. Note that changing the polarization angle also changes the
depth of $V_2$, and therefore the corresponding hopping amplitude
$t_2$. Hence,
systems of atoms where state selective optical potentials can be
implemented
may provide ideal test beds for studying
properties of the two species bosonic Hubbard model with variable
hopping amplitudes and interspecies interaction strength.

Several theoretical works have discussed realizations of novel
phases in the two-species system in an optical lattice
\cite{demler1,duan1,kuklov1,chen1}. Because of the inter-species
interaction, the Mott phase is divided into regions with different
long range orders. These phases can be described in terms of
isospin, a quantum number which describes the occupation state of
a single site by two components \cite{demler1,duan1,kuklov1}. For
a total occupation $2n_0-1$, the states $\left|n_0,n_0-1\right>$
and $\left|n_0-1,n_0\right>$ correspond to isospin states with
$S_z=1/2$ and $S_z=-1/2$ respectively. However, at the superfluid
transition point, which can be approached by decreasing the
strength of the optical lattice, the isospin description breaks
down because of strong density fluctuations. The isospin quantum
number $S_z$ , which is given by the difference in quantized
occupation numbers of the two boson species, becomes ill-defined
at this point. Nevertheless, one can still investigate the effect
of such isospin order in the Mott state on the
superfluid-insulator (SI) transition. This is the key issue that
we are going to address in this work. We note that although there
have also been earlier studies of the SI transitions from such
isospin symmetry broken Mott states \cite{chen1,demler1},
the phase diagram of the system for the entire parameter range
has, to the best of our knowledge, not been charted out
and large parts remain to be explored.

Keeping the above-mentioned experimental and theoretical
scenarios in mind, we shall study a two-species bosonic Hubbard
model described by the Bose-Hubbard Hamiltonian
\cite{demler1,duan1,kuklov1}
\begin{eqnarray}
{\mathcal H} &=& \sum_{\alpha}\left[  \sum_{\left<ij\right>}\left(
-t_{\alpha} b_{i \alpha}^{\dagger} b_{j \alpha} + {\rm h.c}
\right) - \mu \sum_i n_{i \alpha} \right] \nonumber\\
&& + \frac{U}{2} \left[ \sum_{i\alpha}  n_{i \alpha}
(n_{i\alpha}-1) + 2 \lambda
 \sum_i n_{i 1} n_{i 2} \right],\label{Hubbard}
\end{eqnarray}
where $\alpha=1,2$ is the species index, $t_{1(2)}$ denote hopping
amplitudes for the two species between nearest neighbor sites
$\left<ij\right>$, the matrix element $U$ denotes on-site
intra-species Hubbard interaction, $U' =\lambda U$ is the
inter-species interaction and we have taken the chemical potential
$\mu$ to be the same for both the species. Note that a change
in
optical lattice depth by tuning the laser polarization also leads to
a relative shift of chemical potential of the two species. However
this shift is usually small and can always be compensated by
applying an external magnetic field since the two species have
different magnetic moments. For future convenience, we introduce the
ratio $\eta = t_{2}/t_{1}$ and shall take $t_2 \le t_1$ ($\eta \le
1$). Our aim is to study the different phases of the system as a
function of $\mu, \lambda, t_1 \,{\rm and} \, \eta$ for odd total
filling factor. Also, we shall refer to the species index $\alpha$
as the isospin label for the bosons with isospin $S=1/2$.

Before proceeding with the analysis, we summarize the key results of
this work. First, we find that the superfluid-insulator transition
in systems described by Eq.\ \ref{Hubbard} can take place with a)
simultaneous onset of superfludity of species $1$ and $2$ (SF-SF
phase) or b) coexistence of Mott insulating phase of species $2$ and
superfluid phase of species $1$ (${\rm MI}_2+{\rm SF}_1$ phase) or
c), in the case of a unit filling Mott state, depopulation of
species $2$ (a-SF phase).  Second, for a large region of the phase
diagram the superfluid-insulator transition occurs with a
discontinuous jump in the number of
each
species and is therefore first order. Third, there is a second
order quantum phase transition between the a-SF and the SF-SF
superfluid phases which can be viewed as a $n_0=0$ Mott
insulator-superfluid transition for the bosons of species $2$.
Finally our analysis explicitly demonstrates the necessity of
including effects of ${\rm O}(t^2/U^2)$ quantum fluctuations
(beyond the ${\rm O}(t/U)$ mean-field theory) for a correct
quantitative description of the phase diagram and the nature of
the phase transitions in the system.

The organization of the paper is as follows. To put this work in
perspective, we review the results on the Mott phases of the
Hamiltonian in Eq.~(\ref{Hubbard}) in Sec.~\ref{overview}. In
section~\ref{sitrans}, we study the SI transition using ${\rm
O}(t/U)$ mean-field theory and also discuss the shortcomings of such
a theory in the present case. This is followed by
Sec.~\ref{cantrans}, where we implement a canonical transformation
method which takes into account the effect of quantum fluctuation to
${\rm O}(t^2/U^2)$ on the transition and present a detailed phase
diagram of the model. This is supplemented by quantum Monte Carlo
simulations in Section~\ref{mc}. In Section~\ref{sec:detection} we
discuss how the different phases can be detected experimentally.
This is followed by a summary of our results in
Section~\ref{conclusion}.

\section{Review of Mott phases}
\label{overview}
\begin{figure}[t]
\vspace*{-0.7cm}
\centerline{\psfig{file=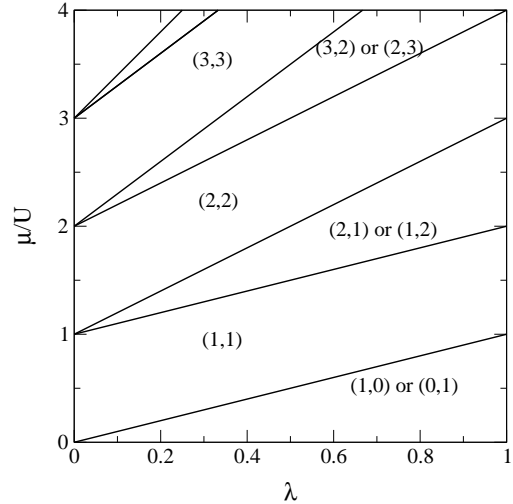,width=0.95\linewidth,angle=0}}
\caption{Schematic phase diagram of two-species boson model in the
Mott insulating state for $t_1=t_2=0$. Notice the two-fold
degeneracy at each site for odd fillings.} \label{fig1}
\end{figure}

In this section, we review the Mott phases of the two species
Bose-Hubbard model \cite{demler1,duan1,kuklov1}. Deep inside the
Mott phase, for $t_1=t_2=0$, the Hubbard Hamiltonian (Eq.\
\ref{Hubbard}) reduces to sum of on-site terms $H_i$ given by
\begin{eqnarray}
H_i &=& -\mu \sum_{\alpha} n_{i\alpha} +\frac{U}{2} \Bigg[
\sum_{\alpha} n_{i \alpha} (n_{i\alpha}-1) +
2 \lambda n_{i 1} n_{i 2} \Bigg]. \nonumber\\
 \label{onsite}
\end{eqnarray}
The phases of $H_i$ are characterized by the ground state of the
system having an integer number of bosons $n_{1,2}(\mu/U, \lambda)$
per site. The phase diagram is shown in Fig.\ \ref{fig1}. Apart from
the usual even filling phases where $n_1=n_2$,  phases with odd
filling $n_1-n_2=\pm 1$, which has no counterpart in single species
systems, occur. For the rest of this paper, we shall concentrate on
phases with odd total filling, where each site is doubly degenerate
($n_1-n_2=\pm 1$) leading to $2^N$ degenerate ground states for a
system with $N$ sites for $t_1=t_2=0$.
\noindent

At finite hopping strengths, this degeneracy is lifted by quantum
fluctuations which can be studied by second order perturbation
theory. More precisely, one can carry out
 this perturbation theory in the regime $U,U',|U-U'|
>> t_1,t_2$ where we are far away from both the SU(2) symmetric
limit ($\lambda=1$) and the vanishing inter-species interaction
limit ($\lambda \ll 1$). In both these limits
perturbation theory breaks down.

To compute the fluctuation correction for a Mott state with an odd
number $n_0$ of atoms on each site, we divide the system into A and B
sublattices (to allow for the possibility of an antiferromagnetic
phase) and use a trial wave-function
$$\left|\Psi\right>=\prod_{i\in A}\prod_{j\in
B}\left|\psi_A\right>_i\left|\psi_B\right>_j$$ where
$$\left|\psi_{A,B}\right>=\cos\frac{\theta_{A,B}}{2}\left|n_0,n_0-1\right>
+e^{i\phi_{A,B}}\sin \frac{\theta_{A,B}}{2}\left|n_0-1,n_0\right>$$
where $\left|n_1,n_2\right>_i$ denotes $n_1$ and $n_2$ atoms of
species $1$ and $2$ at site $i$. A perturbative calculation
yields the ${\rm O}(t^2/U^2)$ correction to the ground-state energy as a
function of the angles $\theta_{A,B}$ and $\phi_{A,B}$:
\begin{eqnarray}
E_f &=& -\frac{Nzt_1^2}{2U} \Bigg[(1+\eta^2)n_0^2(1+\cos\theta_A
\cos\theta_B) \nonumber\\
&& + (1-\eta^2)n_0 (\cos \theta_A+
\cos\theta_B) \nonumber \\ &&+ (1+\eta^2)\left[1-\cos\theta_A
\cos\theta_B\right] \left[ \frac{n_0^2}{2\lambda} +
\frac{n_0^2-1}{2-\lambda}\right] \nonumber\\
&+& \sin(\theta_A)
\sin(\theta_B)\cos(\phi_A-\phi_B) \frac{\eta n_0^2}{\lambda} \Bigg],
\label{efluc}
\end{eqnarray}
where $z$ is the coordination number of the lattice. Minimizing the
\begin{figure}[t]
\centerline{\psfig{file=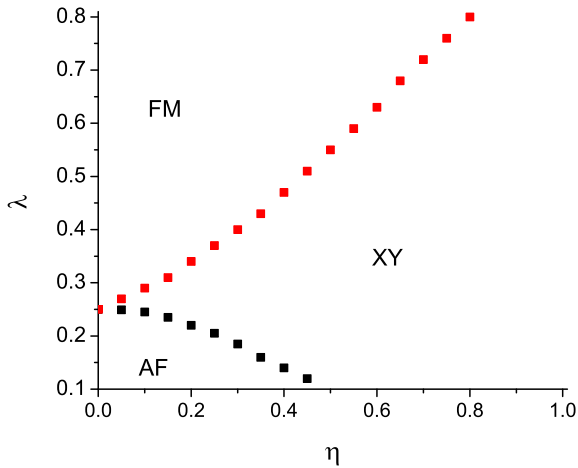,width=
\linewidth,angle=0}} \caption{(Color online) Mean-field phase diagram of the
effective low energy Hamiltonian in Eq.\ \ref{efluc} obtained using
perturbation theory at odd filling for $n_0 = 1$. The phase diagrams
for other values of odd $n_0$ are qualitatively similar in the Mott
insulating regime.} \label{fig2}
\end{figure}
$E_f$ with respect to $\theta_{A,B}$ and $\phi_{A,B}$, we obtain the
phase diagram shown in Fig.\ \ref{fig2} for $n_0=1$ as a function of
the parameters $\eta$ and $\lambda$.  This phase diagram illustrates
presence of three types of phases \cite{duan1}: a) antiferromagnetic
(AF) phase with $\theta_{A(B)}=0,\theta_{B(A)}=\pi$, b)
ferromagnetic (FM) phase with $\theta_A=\theta_B=0$, and c) XY phases
with $\theta_A=\theta_B \ne 0$. The nature of the transitions between
these effective isospin phases can be understood by plotting the
values of the angles $\theta_A$ and $\theta_B$ across the different
phase boundaries. These plots are shown in Fig.\ \ref{fig3}. 
Here the angles $\theta_{A}$ and $\theta_{B}$ have been plotted for three
different values of $\eta$. In the XY and FM phases $\theta_A=\theta_B$ 
and the lines are indistinguishable, whereas in the AF phase one phase takes on a value $\pi$ and the other
$0$. 
We find
that there is always an abrupt jump from the XY to the AF phase across
the AF phase boundary, suggesting that AF-XY transition is first
order. We do not find any canted AF phases. The situation here is
analogous
to the first order melting transition of hard-core bosons
with next nearest-neighbor interaction at half filling \cite{bs}.  The
FM-XY transition, on the other hand, is continuous and proceeds via
continuous change of $\theta_A$ and $\theta_B$.

The phase diagram obtained here agrees qualitatively with that of
Ref.\ \onlinecite{duan1}, although there is a quantitative
difference. In our phase diagram, the tricritical point where all the
phases meet is at $\lambda=0.25, \eta=0$ instead of $\lambda=0.5,
\eta=0$, as found in Ref.\ \onlinecite{duan1}. To understand why this
difference arises, we now map the boson Hamiltonian to an effective
low-energy spin-model. Defining the isospin operators
$S^z_{i}=(n_{1i}-n_{2i})/2$,
$S^x_{i}= (b_{1i}^{\dagger} b_{2i}+b_{2i}^{\dagger} b_{1i})/2$
and
$S^y_{i}= i(b_{1i}^{\dagger} b_{2i}-b_{2i}^{\dagger} b_{2i})/2$
one obtains an effective XXZ model in a
magnetic field \cite{demler1,duan1}
\begin{eqnarray}
{\mathcal H}_{\rm XXZ} &=& -\sum_{\left<ij\right>} \Big[ J_{\perp}
  \left(S^x_{i} S^x_{j}+S^y_{i} S^y_{j} \right) + J_z S^z_{i} S^z_{j}
  \Big] \nonumber \\
&& - B \sum_i S^z_{i} + U(1-\lambda)\sum_i S_{zi}^{2}.
\label{spinmodel}
\end{eqnarray}
The exchange couplings $J_{\perp}, J_z$ and the magnetic field $B$ are
given by
\begin{eqnarray}
J_{\perp}&=& \frac{4t_1^2 \eta n_0^2}{\lambda U}, \quad \quad
B=\frac{2zt_1^2(1-\eta^2)n_0}{U},\nonumber\\ J_z &=& \frac{4t_1^2
}{U}(1+\eta^2) \left[ n_0^2 \left(1-\frac{1}{2\lambda}\right)
-\frac{n_0^2-1}{2-\lambda}\right].
\label{spincoeff}
\end{eqnarray}
Note that for $n_0=1$, AF, FM and XY phases meet when
$J_z=J_{\perp}=0$ at $\lambda=0.5, \eta=0$ provided one neglects the
magnetic field term, as done in Ref.\ \onlinecite{duan1}. However, if
one retains the magnetic field term, the AF and the FM phases will
meet when
$J_z+2B/z=0$ and $J_{\perp}=0$ or $\lambda=0.25, \eta=0$.

The XY phase obtained here is identical to the superfluid
counterflow (SCF) phase obtained in Ref.\ \onlinecite{kuklov1} and
also to the $\nu=1$ bilayer quantum Hall state for small layer
separation where the layer index plays the role of isospin
\cite{steve1,steve2}. The stiffness energy locking the two order
parameter phases together in the XY phases
 can be obtained from Eq.\
\ref{efluc}
\begin{eqnarray}
\rho_s &=& \left(\frac{\partial^2 E_f}{\partial
(\phi_A-\phi_B)^2}\right)_{\phi_A=\phi_B} = \frac{N z t_1^2\eta
n_0^2}{2U \lambda}\sin{\theta_A} \sin{\theta_B} \nonumber
\label{stiff}
\end{eqnarray}

Note that the $U (1-\lambda)\sum_i S_{zi}^{2}$ term in ${\mathcal
H}_{\rm XXZ}$ is a constant since
$S_{zi}^{2}=1/4$
for all the states in the low energy manifold with
$S_{zi}=\pm 1/2$.
Hence this term
does not contribute to the low energy effective Hamiltonian and
does not play a role in the quantum disordering of the XY phase.
The disordering of the XY phase due to quantum fluctuation depends
only on the exchange constants $J_{z}, J_{\parallel}$ and $B$.

\begin{figure}[t]
\centerline{\psfig{file=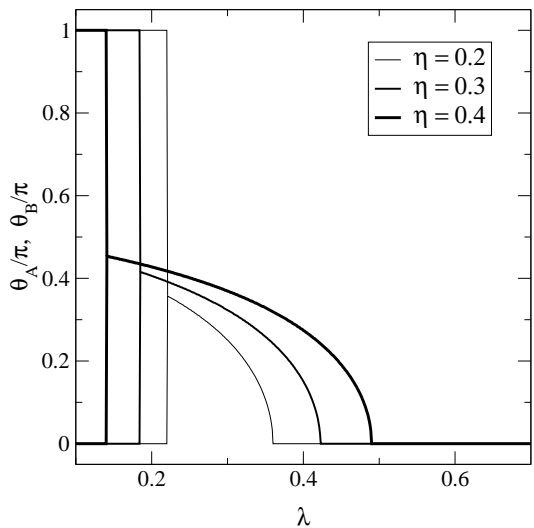,width=1\linewidth,angle=0}}
\caption{Plot of optimum values of $\theta_A$ and
$\theta_B$ as a function of $\lambda$ for different values of
$\eta$. 
In the XY and FM phases $\theta_A=\theta_B$ and the lines for $\theta_{A(B)}$ are indistinguishable whereas
in the AF phase one angle takes on a value $\pi$ while the other takes on a zero value.
We find a discontinuous change in both $\theta_A$ and
$\theta_B$ as the AFM phase is entered, signaling a first-order
transition.} \label{fig3}
\end{figure}
\noindent

\section{${\rm O}(t/U)$ Mean-field theory for the SI transition }

\label{sitrans}

In this section, we shall study the SI transition within ${\rm
O}(t/U)$ mean-field theory by constructing a site factorizable
variational wavefunction which provides an analytical albeit
qualitative understanding of the transition. We shall work in the
parameter regime where $U,U',|U-U'| \gg t_1,t_2$. For the sake of
clarity, although we shall qualitatively comment on the general case,
all calculations in this section from here on shall be performed for
two spatial dimensions and $n_0=1$.

Before carrying
out
the mean-field analysis, we review earlier studies of SI
transition for the two species Bose-Hubbard model
(Eq.~\ref{Hubbard}). In Ref.~\onlinecite{chen1}, the SI transition
has been studied for the case $t_1=t_2$ but with different
inter-species and intra-species interaction strengths and chemical
potentials. This has been done using a standard mean-field
theory~\cite{Sachdev1} corresponding to decoupling the hopping
between sites by introducing order parameter fields
$\Delta_\alpha$, {\emph i.e.}
$b^\dagger_{i\alpha}b_{j\alpha}\approx
b^\dagger_{i\alpha}\Delta_\alpha
+b_{j\alpha}\Delta_\alpha^*-|\Delta_\alpha|^2 $ where the fields
$\Delta_\alpha$ satisfy the self consistency relations
$\Delta_\alpha=\left<b_\alpha\right>$. Their analysis led to the
prediction of three different phases (1) Both species superfluid
(SF-SF). (2) Species 1 superfluid and species 2 in a Mott state
(SF-MI). (3) Species 2 superfluid and species 1 in a Mott state.
(MI-SF). It has been found (erroneously, as we shall see) in
Ref.~\onlinecite{chen1} that the Mott states are always
destabilized by MI-SF or SF-MI phases and there is no direct
transition from the Mott to the SF-SF phase. The transitions are
concluded to be second order as in
the standard single species Bose-Hubbard model \cite{Sachdev1}.
The question of the interplay between the exchange effects and the
SI transition in the region of small $\lambda$ was studied by
Demler {\it et al.}~\cite{demler1} for unit filling factor and
fixed chemical potential
$\mu/U=\frac{1}{2}\lambda$.
Apart from
the phases mentioned above they also found a superfluid phase with
species one superfluid and depopulation of
species $2$.

In our proposed setup, $\lambda$ is not necessarily small  and the
SI transition in this regime has not previously been investigated.
To analyze the SI transition to ${\rm O}(t/U)$ within mean field it
suffices to consider an on-site trial wavefunction \cite{comment1}
\begin{eqnarray}
\left|\Psi_v\right> &=&\prod_i \Bigg(u_0 \left|1,0\right>_i + r
\left|0,1\right>_i + p_1 \left|2,0\right>_i \nonumber \\
&&+ p_2 \left|1,1\right>_i
+p_3 \left|0,2\right>_i +h_1 \left|0,0\right>_i \Bigg) \label{varwave}
\end{eqnarray}
where $u_0$ is the amplitude of the ferromagnetic Mott state
$\left|1,0\right>$ in $\Psi_v$, $h$ and $p$ are amplitudes of
removing, adding bosons of species $1,2$ to the Mott state and $r$ is
the amplitude of isospin-flip. We note here that allowing the
isospin-flip process in the trial wavefunction (Eq.~\ref{varwave}) is
absolutely crucial for correctly taking into account the manifold of
low energy boson states which are degenerate to ${\rm O}(t/U)$. The
normalization of the wave-function yields the constraint $u_0^2 +r^2 +
p_{1}^2 +p_2^2+ p_3^2 +h_{1}^2 =1$. This wavefunction, as we shall
see, is appropriate for studying the SI transition from the FM and the
XY Mott phases. We shall comment about the AFM-SI transition later.

The energy of the variational ground state $
E_v(u_0,r,p_{1},h_{1},p_{2},h_{2}) = \left<\Psi_v\right|{\mathcal
H}\left|\Psi_v\right>$ is given by
\begin{eqnarray}
E_v &=& E_{Mott}+ \Bigg[(p_{1}^2+p_3^2) \delta
E_{1}^{+} + p_2^2 \delta E_2^+ + h_{1}^2 \delta E_{1}^{-} \nonumber\\
&& -\sum_{\alpha=1,2} \frac{\Delta_{\alpha}^2}{zt_{\alpha}} \Bigg],
\label{envar}
\end{eqnarray}
where $E_{Mott}$ is the energy of the Mott state, $z$ is the
coordination number of the lattice, $\delta E_{\alpha}^{\pm}$
denote the energies of adding/removing a boson of species $\alpha$
to/from the Mott state given by
\begin{eqnarray}
\delta E_{1}^{+} &=& -\mu + U , \quad
\delta E_{2}^{+} =  -\mu + \lambda U  \quad
\delta E_{1}^{-} =  \mu ,
\label{spe}
\end{eqnarray}
and the superfluid order parameters $\Delta_{1,2}$ can be
calculated from this variational wave-function:
\begin{eqnarray}
\Delta_1 &=& zt_1 \left<\Psi_v\right|b_{1}\left|\Psi_v\right>=
zt_1 \left(u_0 p_1 \sqrt{2} +r p_2  + u_0 h_1 \right) \nonumber\\
\Delta_2 &=& zt_2 \left<\Psi_v\right|b_{2}\left|\Psi_v\right>=
zt_2 \left(u_0 p_2 + r h_1 + rp_3 \sqrt{2}\right)
\label{orderv}
\end{eqnarray}

Mathematically, it is possible to show that for all of the Mott and
superfluid phases (except the AFM phase for which we need to use two
sublattices), the variational energy has a stationary point. The
parameter values at these points and how they translate into the
various phases is shown in Table~\ref{tab:phasetab}.
\begin{table}[t]
\begin{tabular}{|c|c|c|c|c|c|c|}
  \hline
  Phase & $u_0$ & $r$ & $p_1$ & $p_2$ & $p_3$ & $h_1$ \\
  \hline
   FM Mott-Insulator & 1 & 0 & 0 & 0 & 0 & 0\\
   XY Mott-Insulator &$\neq 0$ &$\neq 0$ & 0& 0&0 & 0\\
   SF$_1$,MI$_2$     &0 & $\neq 0$ & 0 & $\neq 0$ &0 & 0\\
   MI$_1$, SF$_2$    &$\neq 0$ & 0 & 0 & $\neq 0$ &0 &  0\\
   SF$_1$, 0$_2$ (a-SF)     &$\neq 0$ & 0& $\neq 0$ & 0 &0 & $\neq 0$ \\
   SF$_1$,SF$_2$     &$\neq 0$ & $\neq 0$ & $\neq 0$ & $\neq 0$ &$\neq 0$& $\neq 0$\\
  \hline
\end{tabular}
\caption{Parameter values of the variational wavefunction corresponding to
the different phases.\label{tab:phasetab}}
\end{table}
The transition to superfluidity from the Mott state occurs when it
becomes energetically favorable to have non-zero $\Delta_{\alpha}$
${\it i.e.}$ non-zero amplitudes of additional particles and holes
($p$ and $h$) in the variational ground state. For our purposes, it
is sufficient to take all the coefficients real. This amounts to
setting the phase of the superfluid order parameter to zero and does
not affect the variational energy of the state. Note that the
wavefunction (Eq.\ \ref{varwave}) is general enough to incorporate
both the FM ($u_0=1$) and the XY ($u_0=\cos(\theta/2)$
$r=\sin(\theta/2)$) phases of the Mott state. However, since these
two states are degenerate to ${\rm O}(t/U)$, our simple mean-field
treatment cannot distinguish between their isospin order. In
Secs.~\ref{cantrans} and \ref{mc} we shall carry out more
sophisticated treatments of our model which will take into account
the effect of ${\rm O}(t^2/U^2)$ fluctuations using canonical
transformation and quantum Monte-Carlo. In this section, we shall
analyze the {${\rm O}(t/U)$ mean-field theory and point out certain
qualitative features of the phase diagram.

\subsection{General features of the phase diagram}
\label{genf}

Although the variational wave function in this section excludes
second order exchange effects,
 the qualitative features of the SI
transition from the FM and the XY phases for $n_0=1$ can be
understood from Eqs.\ \ref{envar} and \ref{spe}. Consider, for
example, approaching the SI transition from FM/XY Mott phase. For
$\mu \ll U,\lambda U $, since $\delta E_{1}^{-} \ll \delta
E_{1}^{+},\delta E_{2}^{+}$, at the SI transition point $t_1
\equiv t_1^{c} = \delta E_{1}^{-}/z$ and the energy of the
variational wavefunction is minimized with $r=0$, $u_0 \sim 1$,
$p_1=p_2=p_3=0$, $h_1\ne 0$. Consequently, from Eq.~\ref{orderv},
we have $\Delta_2 =0$, $\Delta_1 \ne 0$, and also
\begin{eqnarray}
n_1 &=& \left<\Psi_v\right|b_{1i}^{\dagger}b_{1i}\left|\Psi_v\right>
= \left(u_0^2 + 2p_1^2 +p_2^2\right ) \equiv 1 \nonumber\\
n_2 &=& \left<\Psi_v\right|b_{2i}^{\dagger}b_{2i}\left|\Psi_v\right>
= \left(r^2 + 2p_3^2 +p_2^2\right ) = 0  \label{density1}
\end{eqnarray}
Thus the transition to superfluidity occurs with complete
depopulation of species $2$. We refer to this phase as a-SF.
Alternatively one can view this phase as SF$_1$-MI$_2$ with a zero
filling factor in the Mott phase. Numerically, we find that such a
depopulation occurs till a critical value of $\mu= \mu_{c1}$.


\begin{figure}[t]
\centerline{\psfig{file=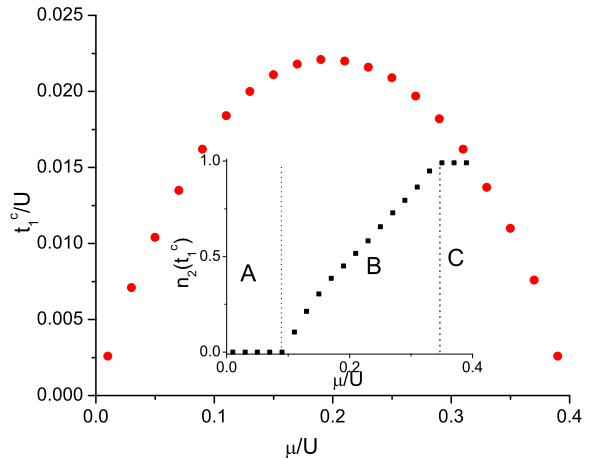,width=\linewidth,angle=0}}
\caption{(Color online) ${\rm O}(t/U)$ mean-field results for the SI transition
from the XY phase
for $\lambda=0.4$ and $\eta=0.8$. The inset shows $n_2$ at the
transition. Superfluidity sets in a) with depopulation bosons of
species $2$ in region A (a-SF phase) b) simultaneously for both
species in region B (SF-SF) and c) with Mott phase for $2$ and
superfluid for $1$ in region C (${\rm MI}_2+{\rm SF}_1$). The vertical
dotted lines are
guides
to the eye and represent the positions of $\mu_{c1}$ and
$\mu_{c2}$ (see text).} \label{figphasexy}
\end{figure}

\noindent


In the other limit, when $\mu \simeq \lambda U \ge \mu_{c2}$, it is
much more favorable to destabilize the Mott state by adding a particle
of species 2 since $\delta E_2^+ \ll \delta E_1^-,\delta E_1^+$. As a
result the transition occurs with $u_0=0$, $r \simeq 1$ and $p_2 \ne
0$. Consequently, the transition takes place with
\begin{eqnarray}
\Delta_2 &=& 0, \quad n_2,\,\, \Delta_1, \,\, n_1 \ne 0
\end{eqnarray}
$i.e.$, we have a transition which is accompanied by a jump of
population species $2$ at the transition. The phase consists of a Mott
insulator of species $2$ (since $\Delta_2=0$) and superfluid of
species $1$. We call this state ${\rm MI}_2+{\rm SF}_1$.

For $\mu_{c2} > \mu > \mu_{c1}$,  $\delta E_2^{+}$ and
$\delta E_1^{-}$ are comparable and the energy of the ground state at the
transition is minimized for $r,u_0 \ne 0$ and $p_2 \ne 0$. In this
case, provided $\eta \ne 0$, both $\Delta_1$ and $\Delta_2$ are
non-zero at the transition implying simultaneous
onset
 of superfluidity of species $1$ and $2$ (referred to as
SF-SF). The width of this region is expected to be large at large
$\eta$, since higher $t_2$ makes it energetically more favorable
to realize superfluidity of species $2$.

The values of $\mu_{c1}$ and $\mu_{c2}$ are shown for representative
values of $\eta=0.8$ and $\lambda=0.4$ in Fig.\ \ref{figphasexy}.
The phase diagram corroborates the above discussion. From the inset
of Fig.\ \ref{figphasexy}, we find that there are three distinct
regions
 where the SI transition takes place with a) depopulation
of species $2$ (region A), b) simultaneous setting of
superfluidity of the two species (region B), and c) Mott
insulating phase of species $2$ and superfluidiy of species $1$
(region C). The situation here is in sharp contrast to the even
filling case which will always have an intermediate state with
superfluidity of species $1$ and insulating state of species $2$
for $0 < \eta < 1$.

Upon further increase of $t_1$ from the critical value $t_{1c}$, two
scenarios are possible. If superfluidity sets in with
depopulation, increasing $t_1$ does not change the situation
further. On the other hand, if the transition occurs to either
the
 SF-SF or ${\rm MI}_2+{\rm SF}_1$ phase, upon increasing $t_1$,
the fraction of B atoms in the superfluid decreases as shown in
Fig.\ \ref{orderfm} for a set of representative values of $\eta$,
$\lambda$ and $\mu$. Finally, one crosses a critical value
$t_1^{\ast}$ at which it becomes energetically favorable for the
system to depopulate. This happens at large enough $t_1 \ge
t_1^{\ast} \sim \delta E_1^{+}/z$, at which the variational energy
minima shifts to $u_0 \ne 0$, $r=p_2=0$. Beyond this point, we
only find superfluidity of species $1$. Within ${\rm O}(t/U)$
mean-field theory, such a transition from SF-SF to a-SF phase is
found to be first order since $\sum_{\alpha=1,2}
\Delta_{\alpha}^2/zt_{\alpha}$ is discontinuous across the
transition.

Although we do not show it explicitly here, a similar consideration
remains valid for the SI transition from the AFM phase. This can again
be seen by dividing the lattice into the usual $A$ and $B$ sublattices
and constructing an appropriate two sublattice variational
wavefunction. We do not find any translational symmetry broken
superfluid phases. We also note that the above discussions
have to be modified for
$n_0 \ne 1$, where the Mott state can also be destabilized by
adding holes of species $2$. For example, when $n_0 \gg 1$, for
$\mu \ll \lambda$, $\delta E_{1}^{-} \approx \delta E_{2}^{-}$.
Thus if $\eta \ne 0$, we expect the ground state energy to be
always minimized for $u_0,r \ne 0$ at the transition. Consequently,
there will be no depopulation for any finite $\eta$ in this limit.
In the rest of this work, we shall restrict all discussion to odd
fillings with $n_0=1$.


\begin{figure}[t]
\centerline{\psfig{file=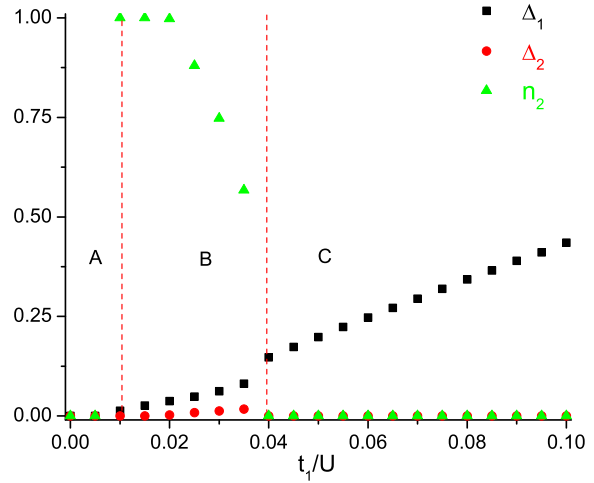,width=\linewidth,angle=0}}
\caption{(Color online) A plot of the order parameters $\Delta_1$, $\Delta_2$, and
$n_2$ for $\mu/U=0.57$, $\lambda=0.6$ and $\eta=0.2$ as a function of
the hopping amplitude $t_1$. The system enters the ${\rm MI}_2+{\rm
SF}_1$ phase at $t=t_1^c$ (vertical dotted line between regions A and
B) from a FM Mott phase (region A).  Note that the transition occurs
with a spontaneous jump of $n_2$ and is hence expected to be first
order.  As we increase $t_1$, both species becomes superfluid (region
B), until $t_1$ reaches $t_1^{\ast}$ (vertical dotted line between
region B and C) where the system depopulates. The depopulation occurs
with a jump in $\Delta_1$ and is therefore first order.}
\label{orderfm}
\end{figure}

\noindent

\subsection{Necessity of going beyond the ${\rm O}(t/U)$ mean-field theory}
\label{beyondmft}

We now discuss the limitations of the mean-field theory to set the
stage for incorporating the fluctuation effects. To understand why
using the mean-field theory is dangerous in the present context,
consider plotting the mean-field phase diagram at a fixed
$\mu/U =0.5$ and $t_1/U=0.04$
as a function of $\eta$ and $\lambda$.
Such a phase diagram is shown in Fig.~\ref{figmft04}. Here, we have used the
phase diagram (Fig.~\ref{fig2}) of the XXZ model [Eq.\ (\ref{spinmodel})]
to determine the isospin phases  since the  ${\rm O}(t/U)$
mean-field theory can not distinguish between them. As can be seen,
the phase diagram (Fig.\ \ref{figmft04}) corroborates the expectations based
on the qualitative discussion of the Sec.\ \ref{genf}. For small $\eta$
(transition from FM/AFM phases), the system favors a-SF phase while
for larger $\eta$ (transition from the XY-phase) the SF-SF phase dominates.

\begin{figure}[t]
\centerline{\psfig{file=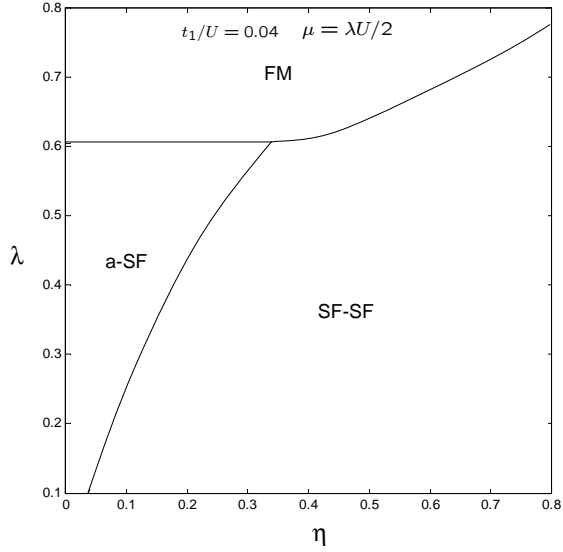,width=0.9\linewidth,angle=0}}
\caption{ ${\rm O}(t/U)$ mean-field phase diagram for the two-species
model as a function $\eta$ and $\lambda$ for
$\mu/U=\lambda/2$
and $t_1/U=0.04$. In the absence of second order fluctuation corrections, the SF-SF
phase borders to the FM phase implying a discontinuous change in the
population of species 2 for a large region of parameter phase.}\label{figmft04}
\end{figure}

\noindent

However, consider now plotting such a phase diagram near $\mu_{c1}$
or $\mu_{c2}$.  Clearly, we expect that incorporating exchange
effects will make it harder for the isospins to flip, since now it
costs an energy ${\rm O}(t^2/U^2)$. This will, in general, shift the
positions of $\mu_{c1}$ and $\mu_{c2}$ from their mean-field values.
Therefore, near $\mu_{c1}$ or $\mu_{c2}$, the phase diagrams in the
$\eta-\lambda$ plane predicted by the mean-field theory will be
qualitatively different from the true phase diagrams. In this sense,
the failure of the ${\rm O}(t/U)$ mean-field theory in the present
case is much more severe compared to the usual SI transition for
single species bosons. However, as long as we are away form the
critical $\mu$ values, such a mean-field theory gives qualitatively
correct results and therefore the scenario described in the previous
section remains largely valid.

Another problem of the ${\rm O}(t/U)$ mean-field theory is that it
overestimates the jump of $n_1$ or $n_2$ at the transition since it
does not take into account the energy cost of an isospin flip.
Consequently, it can erroneously predict first-order MI-SF or a-SF
to SF-SF transitions where in reality such transitions might be
second order. Also, as we shall see in the next section, the shapes
of the transition curves and topology of phase boundaries change
quite a bit upon inclusion of the fluctuation corrections.

Thus, although the ${\rm O}(t/U)$ mean-field theory correctly
predicts the qualitative nature of the MI-SF transition for most
parts of the phase diagram
 it fails drastically either when we are close to
$\mu_{c1}$ or $\mu_{c2}$ or when we want to estimate the order of
the transition. In the next section, we remedy this failure by
incorporating the ${\rm O}(t^2/U^2)$ fluctuation corrections.

\section{Canonical Transformation}
\label{cantrans}

The effect of fluctuation to second order in $t/U$ can be taken into
account using a suitable canonical transformation method. We
describe the implementation of this method in subsection \
\ref{implementaion} and present the phase diagrams in subsection\
\ref{phasedia}.

\subsection{Implementing the canonical transformation}
\label{implementaion} We begin by separating the Bose-Hubbard
Hamiltonian (Eq.\ \ref{Hubbard}) into an onsite term $H_0 = \sum_i
H_i$ (Eq.\ \ref{onsite}) and the hopping terms $T$. The first step
is to write $T$ in terms of sum over bonds $\sigma$ between the
neighboring lattice site. To this end, as shown in Fig.\
\ref{bondfig}, we can decompose the hopping into hopping on vertical
and horizontal bonds, labeled $\sigma_{v,h}$, between adjacent
sites. The hopping term can then be written as a sum over bonds
\begin{eqnarray}
T &=& \sum_{\sigma} T_{\sigma} = \sum_{\sigma_h} T_{\sigma_h} +
\sum_{\sigma_v} T_{\sigma_v} \label{kin1} \\
T_{\sigma_h} &=& -\sum_\alpha t_\alpha\left(b_{\sigma_{h_R} \alpha}^{\dagger}
b_{\sigma_{h_L} \alpha} + {\rm h.c}\right) \label{kin2} \\
T_{\sigma_v} &=& -\sum_\alpha t_\alpha\left(b_{\sigma_{v_U} \alpha}^{\dagger}
b_{\sigma_{v_D} \alpha} + {\rm h.c}\right) \label{kin3}
\end{eqnarray}

\begin{figure}[t]
\centerline{\psfig{file=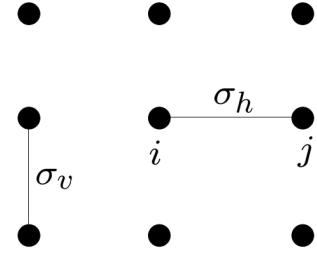,width=0.5\linewidth,angle=0}}
\caption{Bonds in a 2D square lattice. There are two types of bonds,
horizontal and vertical labeled by $\sigma_h$ and $\sigma_v$
respectively. The horizontal bond shown is denoted by $\sigma_{h}$.
The sites on the left and right sides of this bond of (sites $i$ and
$j$ in this case) are labeled by $\sigma_{h_L}$ and $\sigma_{h_R}$
respectively.} \label{bondfig}
\end{figure}
\noindent

We now seek a unitary transformation that will capture the effects
of the second order exchange effects. We shall only consider the
case $n_0=1$, although generalization to other values of $n_0$ is
straightforward. To do this, we first introduce the projection
operators $P_\sigma$ acting on the two sites associated with each bond $\sigma$.
$P_{\sigma}$ projects the state of the system to the manifold of states having one
particle on each site of the bond. Such a projection operator can be
decomposed into two parts depending on whether the bosons occupying
the sites of the bond are of the same or different species:
$P_\sigma=P_\sigma^0+P_\sigma^1$.  For instance for a horizontal bond we can write
\begin{eqnarray}
P_{\sigma_h} &=& P_{\sigma_h}^0 + P_{\sigma_h}^1 \nonumber\\
P_{\sigma_h}^0 &=& \left(\left|1,0\right>\left<1,0\right|\right)_{\sigma_{h_L}}
\otimes\left(\left|1,0\right>\left<1,0\right|\right)_{\sigma_{h_R}} \nonumber\\
&& +\left(\left|0,1\right>\left<0,1\right|\right)_{\sigma_{h_L}}
\otimes\left(\left|0,1\right>\left<0,1\right|\right)_{\sigma_{h_R}}
\label{pro1}\\
P_{\sigma_h}^1 &=& \left(\left|0,1\right>\left<0,1\right|\right)_{\sigma_{h_L}}
\otimes\left(\left|1,0\right>\left<1,0\right|\right)_{\sigma_{h_R}} \nonumber\\
&& + \left(\left|1,0\right>\left<1,0\right|\right)_{\sigma_{h_L}}
\otimes\left(\left|0,1\right>\left<0,1\right|\right)_{\sigma_{h_R}}
\label{pro2}
\end{eqnarray}
where the state $\left|1,0\right>$ denotes
$\left|n_1=1,n_2=0\right>$ as before and the subscript on the
parentheses denotes the bond and which of the sites the operators
act on
(cf.~Fig.~\ref{bondfig}). Using the projection operator
(Eqs.~\ref{pro1},\ref{pro2}), we now decompose $T_\sigma$ (Eq.\
\ref{kin1}) into two parts
\begin{eqnarray}
T_\sigma &=& P_\sigma^\bot T_\sigma P_\sigma^\bot+
\left(T_\sigma P_\sigma+P_\sigma T_\sigma\right)=T_\sigma^0+T_\sigma^1
\label{toperator}
\end{eqnarray}
where $P_\sigma^\bot=1-P_\sigma$. The idea is now to use these results to
seek a unitary transformation
\begin{eqnarray}
H^*&=& e^{iS}{\mathcal H}e^{-iS}=
{\mathcal H}+[iS,{\mathcal H}]+\frac{1}{2}\left[iS,\left[iS,
{\mathcal H}\right]\right]+...\nonumber\\
\label{trans}
\end{eqnarray}
which eliminates the terms $T_\sigma^1$ to ${\rm O}(t/U)$. It turns out that
a suitable choice of $S$ is
\begin{eqnarray}
S &=& \frac{i}{\lambda U}\sum_\sigma [\lambda P_\sigma^0+P_\sigma^1,
T_\sigma] = \frac{i}{\lambda U}\sum_\sigma [P_\sigma^{\lambda},
T_\sigma]
\label{transop}
\end{eqnarray}
where we have introduced the notation $P_{\sigma}^{\lambda}= \lambda
P_{\sigma}^0 + P_{\sigma}^1$ for future convenience. We now expand $H$
(Eq.\ \ref{trans}) in powers of $S$. Since $S$ is first order in
$t/U$, this is equivalent to an expansion in $t/U$
and we have to ${\rm O}(t^2/U^2)$,
\begin{eqnarray}
H^*&=& H_0+\sum_\sigma T_\sigma +\left[iS,H_0+\sum_\sigma
T_\sigma \right]\nonumber\\
&& +\frac{1}{2}\left[iS,\left[iS,H_0\right]\right] +{\rm O}(t^3/U^3)
\label{expansion}
\end{eqnarray}
We now evaluate the different terms in Eq.~(\ref{expansion}). The
algebra is straightforward, but lengthy and we present some details
in Appendix A. The final result, to ${\rm O}(t^2/U^2)$, is
\begin{eqnarray}
H^*&=&H_0+\sum_\sigma P_{\sigma}^{\bot} T_{\sigma} P_{\sigma}^{\bot} \nonumber\\
&& -\frac{1}{2\lambda U}\sum_{\sigma}\left[P_\sigma^\lambda T_\sigma^2 P_\sigma
- T_{\sigma} P_{\sigma}^{\lambda}T_{\sigma} +{\rm h.c.}\right]\nonumber\\
&& -\frac{1}{\lambda U}\sum_{\sigma,j} \left[ P_{\sigma}^{\lambda}
T_{\sigma} T_{\sigma+j} - T_{\sigma} P_{\sigma}^{\lambda} T_{\sigma+j}
\right. \nonumber\\
&& \left. - 2 \left(P_{\sigma}^{\lambda} T_{\sigma}T_{\sigma+j}P_{\sigma+j}
- T_{\sigma} P_{\sigma}^{\lambda}P_{\sigma+j} T_{\sigma+j} \right)
+{\rm h.c.}\right] \nonumber\\
\label{hamf}
\end{eqnarray}
where the sum over $j$ extends over bonds which are nearest neighbors
to $\sigma$. We note that the third and the fourth terms of Eq.\
\ref{hamf}
represent
the effective XXZ model [Eq.\ (\ref{spinmodel})] of
Section~\ref{overview} and the two-particle hopping processes
respectively, whereas the terms in the last line
involve hopping operators on neighboring bonds and are expected to be important in
the superfluid phases.

\begin{figure}[t]
\centerline{\psfig{file=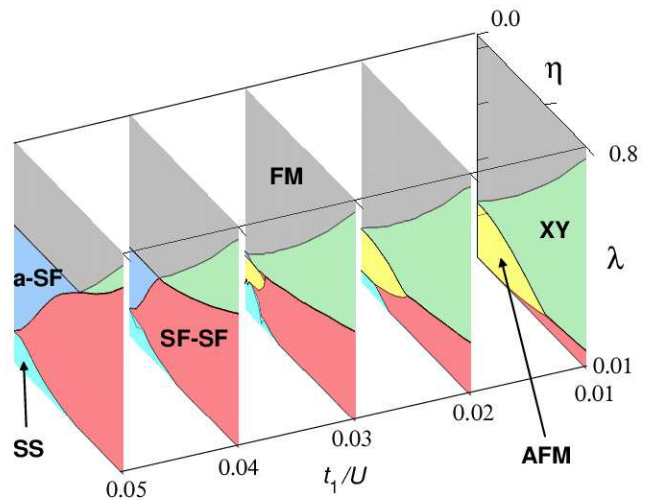,width=1\linewidth,angle=0}}
\caption{(Color online) ${\rm O}(t^2/U^2)$
mean-field phase diagram for the two-species model as a function
$\eta$ and $\lambda$ for $\mu=\lambda U/2$. The plot for
$t_1/U=0.04$ should be compared with the phase diagram obtained to
${\rm O}(t/U)$ in
Fig.~\ref{figmft04} (see text). Note the gradual evolution of the
different phases with increase of $t_1$.
The existence of the multicritical point is due to the special
symmetry at $\mu=\lambda U/2$ where adding a hole or a particle of
species 2 to the FM Mott state cost equal energies.
Quantum Monte Carlo however reveals that these multicritical points
can be split (See Sec.~\ref{mc}). \label{figm5}}
\end{figure}
%

\begin{figure}[t]
\centerline{\psfig{file=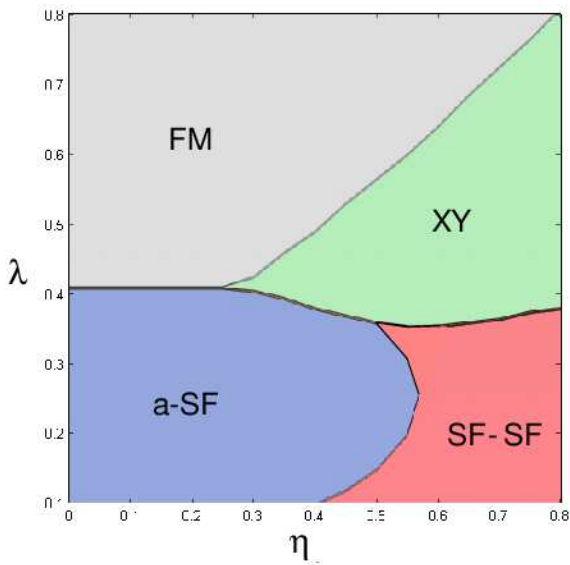,width=0.9\linewidth,angle=0}}
\caption{(Color online) ${\rm O}(t^2/U^2)$ mean-field phase 
diagram for the
two-species model as a function $\eta$ and $\lambda$ for $\mu/U=0.1
\lambda$ and
$t_1/U=0.01$.
The superfluidity is a-SF for a large
parameter regime as predicted by the ${\rm O}(t/U)$ mean-field theory.}
\label{figm1p1}
\end{figure}


\subsection{Phase Diagram}
\label{phasedia}

The phase diagram of $H^{\ast}$ is obtained by dividing the lattice
into two sublattices $A$ and $B$ and using an on-site variational
wavefunction $\left|\psi_v\right>=\prod_{i\in A}\prod_{j\in B}
\left|\psi\right>_i \left|\psi\right>_j$. The division into two
sublattices is essential for taking into account the AFM phase. We
note that this is equivalent to generalizing the  mean-field
treatment of Sec.\ \ref{sitrans} to incorporate second order
fluctuation corrections. Although it is cumbersome to evaluate the
expectation value $\left<\psi_v\right|H^{\ast}\left|\psi_v\right>$
analytically, it can be calculated numerically by representing the
various operators in the Hamiltonian as matrices in an appropriately
chosen basis. The task of minimizing
$\left<\psi_v\right|H^{\ast}\left|\psi_v\right>$ is then a numerical
optimization problem. Truncating the Hilbert space to have at most
two particles on each site, we perform constrained (to keep the norm
to unity) optimization for each point in the phase diagram. We use a
sequential quadratic programming algorithm
from
the MATLAB (TM) optimization toolbox for this task. Due to
nontrivial energy landscapes and possible existence of first order
transitions, several starting points,including random starting
points, were used as input to the algorithm.

First, we show the phase diagram in the $\eta-\lambda$ plane for
$\mu/U = 0.5\lambda$ in Fig.~\ref{figm5}, which shows the gradual
evolution of the phases of the system as $t_1$ is increased. A
comparison of Fig.\ \ref{figm5} for $t_1/U=0.04$ to
its ${\rm O}(t/U)$ mean-field counterpart [Fig.\ (\ref{figmft04})],
immediately shows us the importance of incorporating the exchange
effects. Whereas the ${\rm O}(t/U)$ mean-field phase diagram shows
a large boundary between the FM and the SF-SF phase indicating a
first order transition, Fig\ \ref{figm5} shows only second order
phase boundaries and no direct transition between FM and SF-SF
phases. This clearly points out that incorporating the exchange
effects can lead to qualitatively different results.

The transition between the FM and the a-SF phases
is second order, as expected. The transition between the a-SF and
SF-SF phases is also found to be second order, in contrast to the
prediction of the ${\rm O}(t/U)$ mean-field theory. This is a
consequence of incorporating the second order fluctuation
corrections. We note that the a-SF - SF-SF transition can
alternatively be viewed as a Mott-insulator (with $n_0=0$) -
superfluid transition of species $2$ in the presence of species $1$ in
a superfluid state.  The supersolid (SS) phase obtained for small
values of $\lambda$ and $\eta$ represents a superfluid phase with
broken sublattice symmetry. This is precisely the region where
$zt^2/\lambda U$ becomes large and the perturbation theory breaks
down. We shall see in the next section using Monte Carlo that the SS
phase is indeed an artifact and signifies the breakdown of
perturbation theory.

Similar phase diagrams for $\mu/U=0.1 \lambda$ and $\mu/U=0.9
\lambda$ and
$t_1/U=0.01$
are shown in Figs.\ \ref{figm1p1} and
\ref{figm9p1} respectively. These plots
confirm
 that the qualitative expectations of the ${\rm O}(t/U)$
mean-field theory. For $\mu/U=0.1 \lambda$, we find a large a-SF
region and the transition to a-SF occurs from both FM and XY
phases [Fig.\ (\ref{figm1p1})], the transitions from XY to a-SF
being first order. For $\mu/U=0.9 \lambda$ (Fig.\ \ref{figm9p1}),
the a-SF phase is replaced by ${\rm MI}_2+{\rm SF}_1$. Here we
find a direct first order transition between the FM and ${\rm
MI}_2+{\rm SF}_1$ phases. These first order transitions are in
perfect agreement with the predictions of the ${\rm O}(t/U)$
mean-field theory.


\begin{figure}[t]
\centerline{\psfig{file=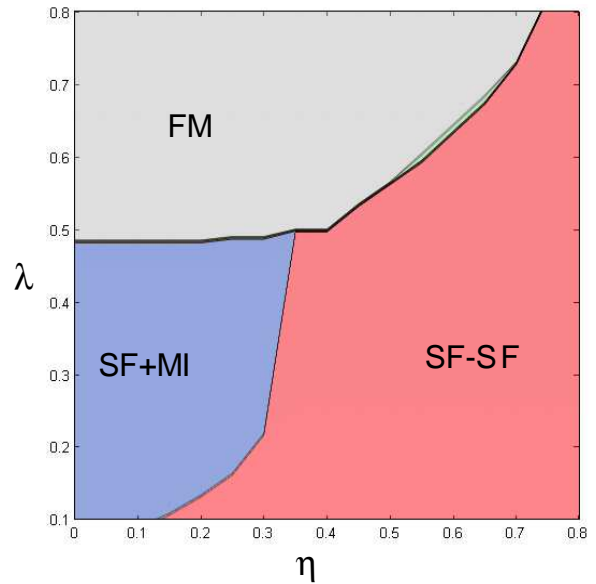,width=0.9\linewidth,angle=0}}
\caption{(Color online) ${\rm O}(t^2/U^2)$ phase diagram for the
two-species model as a function $\eta$ and $\lambda$ for $\mu/U=0.9
\lambda$ for
$t_1/U=0.02$.
We see a large area of ${\rm MI}_2+{\rm
SF}_1$ phase again in accordance with the mean-field theory
prediction.}
\label{figm9p1}
\end{figure}

\noindent


\section{Quantum Monte Carlo}
\label{mc} To verify that the inclusion of ${\rm O}(t^2/U^2)$
corrections using the canonical transformation procedure as carried
out in section~\ref{cantrans} really gives an improvement over the
${\rm O}(t/U)$ mean-field theory in section~\ref{sitrans}, we have
performed quantum Monte Carlo (QMC) studies using the Stochastic
Series Expansion method introduced by Sandwik ${\it et\,al.}$
\cite{Sandwik1,Sandwik2}. Here we have used a particular form of
these updates, directed loop-updates. For details see
Ref.~\onlinecite{Syljuasen}. The basis states used were from a
truncated Hilbert space
in which each site hosts at most two atoms per site,
 in the same
way as done in the mean-field and canonical transformation
treatments. We expect such a truncation to be adequate for
reproducing the phase diagram since we work with $n_0=1$ and are
always close to the MI-SF transition. We have investigated phase
diagrams in the range $\mu/U=0.1\lambda - 0.8\lambda$ and found
Monte-Carlo results agreeing well with the qualitative predictions
of both Secs.\ \ref{sitrans} and \ref{cantrans} and we now turn to
a critical comparison between the results obtained by different
methods.

\begin{figure}[t]
\centerline{\epsfig{file=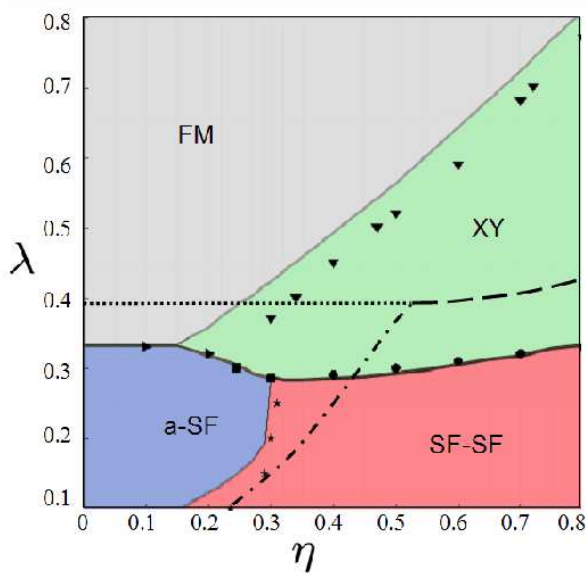,width=0.9\linewidth,angle=0,clip}}
 \caption{(Color online) Comparative phase diagram for the
two-species model as a function $\eta$ and $\lambda$ for
$\mu/U=0.25\lambda$ for
$t_1/U=0.02$.
The discrete points represent the phase
boundaries as calculated using quantum Monte-Carlo while the colored regions
are obtained from the ${\rm O}(t^2/U^2)$ mean field theory. The shape of the 
markers
represent the different phase boundaries; a-SF/SF-SF (stars) ; SF-SF/XY (circles);
a-SF/FM (right triangles); FM/XY (up triangles); a-SF/XY (squares). As comparison 
the
phase boundaries obtained using the ${\rm O}(t/U)$ mean field theory are also shown 
as lines;
a-SF/Mott (dotted); a-SF/SF-SF (dash-dotted); SF-SF/Mott (dashed).
 } \label{figmc1}
\end{figure}

First, QMC confirms our suspicion that the appearance of
the SS phase for small $\lambda$ is indeed an artifact of the
breakdown of the second order perturbation theory at small $\lambda$.
Second, comparisons with QMC show that the
details
of the phase diagram are much better reproduced using the canonical
transformation method.

A comparison between the different methods can be seen in
Fig.~\ref{figmc1} where the phase diagram has been drawn for
$\mu/U=0.25\lambda$, $t_1/U=0.02$. The dotted, dashed and
dash-dotted lines are the phase boundaries as obtained using the
${\rm O}(t/U)$ mean field theory of section~\ref{sitrans}. As can be
seen there is a large discrepancy between the location of the phase
boundaries as compared to the ${\rm O}(t^2/U^2)$ mean field theory.
The discrete set of points represent the phase boundaries obtained
by QMC. Comparing the three methods we thus see that the inclusion
of ${\rm O}(t^2/U^2)$ effects yields the phase boundaries with great
accuracy.

We also find that the Monte Carlo study predicts the a-SF/SF-SF
transition to be second order.  This validates the result obtained
using the canonical transformation method and shows that the
expectation of a first order a-SF-SF-SF transition based on ${\rm
O}(t/U)$ mean-field theory (Sec.~\ref{sitrans}) was an artifact of
omitting ${\rm O}(t^2/U^2)$ corrections.

An interesting prediction, and possible failure, of the ${\rm
O}(t^2/U^2)$ mean-field theory is the existence of continuous
parameter region having points where four phases meet (cf. the
diagrams for $t_1=0.04$ and $t_1=0.05$ in Fig.~\ref{figm5}).
However, using QMC for systems of sizes up to $20\times20$ sites and
inverse temperatures $\beta=1500/U$ for parameter values
$\mu=0.5\lambda U$, $t_1=0.04$, suggests that although the phase
boundaries come very close they do not meet at a single point but a
small region showing a first order transition between the a-SF and
the XY phase seems to remain.

\noindent


\section{Detecting the different phases}
\label{sec:detection}

The traditional way of examining the existence of superfluidity in
trapped boson systems is to switch off the trap, let the cloud of
atoms expand freely and image the expanding cloud. The momentum
distribution of the atoms inside the trap can then be inferred by
looking at their position, or equivalently density, distribution in
the expanded cloud. Since the momentum distribution function of the
atoms is characterized by the presence/absence of coherence peaks in
the superfluid/Mott insulating states, such a measurement serves as
a qualitative probe of the state of the atoms inside the trap
\cite{Bloch1}. In our proposed setup, however, such a simple
expansion alone, which can not distinguish between the two species,
will not be able to distinguish between all the different phases.
Nevertheless, since the two species have different magnetic moments
($m_F = -2$ and $m_F=-1$), it is possible to separate them during
the expansion using a pair of Stern-Gerlach magnets
\cite{stenger,blochnew}.
The expanding cloud will then be separated into two clouds if both
species of atoms are present in the system. This, together with the
momentum distribution measurement, will qualitatively distinguish
between all the phases obtained in this work. To make this statement
concrete, let us consider the phase diagram shown in Fig.\
\ref{figm1p1}. Here the proposed set of measurements will lead to a)
single cloud with no coherence peak (FM phase) or b) single cloud
with coherence peak (a-SF phase), or c) two clouds with no coherence
peaks (XY phase) or d) two clouds with a coherence peak (SF-SF
phase). Thus this methods allows, for instance, the detection of the
mixed phases (${\rm MI}_2+{\rm SF}_1$, a-SF, and SF-SF). It further
provides a tool for finding the phase transitions between the Mott
phases for $n_0=1$. The second order transition from the XY-phase to
the FM phase, for example, will be associated with gradual depletion
of the atoms in one of the clouds whereas the transition to the AFM
phase will be characterized by an abrupt change from a single to a
double cloud.

The above mentioned detection technique, however, does not provide
any evidence of the isospin order in the Mott states since they can
not probe the spatial correlations between the atoms at a lattice
scale. Such correlations can be probed, for example, by tilting the
optical lattice with a potential gradient \cite{Bloch1}. Deep inside
the Mott phase, such a potential gradient will excite the system
only if ${\mathcal E} = E_{dipole}$, where ${\mathcal E}$ is the
potential energy shift between adjacent lattice due to the field
gradient and $E_{dipole}$ is the dipole formation energy
\cite{Bloch1,SKS}. The dipole formation energy will sharply change
across the phase transition lines between AFM-XY and AFM-FM phases
and consequently the peak in the excitation width, measured in Ref.\
\onlinecite{Bloch1} as a function of the applied field gradient,
shall show an abrupt shift at the transitions across these phases.
In contrast, there will be a gradual shift of the peak position as
one moves from the XY to the FM phase. Alternatively, isospin order
can also be measured by probing noise correlation of the expanding
clouds \cite{Han, Bloch3, altman}. For example, a transition from
the FM to AFM isospin states will be marked by appearance of
additional peaks in the noise spectrum at half the reciprocal
lattice vector. The detection of the XY phase can also be obtained
by the Ramsey spectroscopy technique as suggested in Ref.\
\onlinecite{kuklov1}.

Another possible way of detecting the phases is to image the
expanding cloud by passing a linearly polarized laser beam through
it. As shown in Ref.\ \onlinecite{carusotto04}, the angle of
rotation of the plane of polarization ($\theta_{\rm rot}$) of the
outgoing laser beam is proportional to the net $m_z$ along the
direction of propagation ($x_{\perp}$)of the beam: $ \theta_{\rm
rot} \sim \int dx_{\perp} m_z(x_{\perp})$. $\theta_{rot}$ can then
be easily measured by passing the outgoing beam through a crossed
polarizer since the intensity of the beam coming out of the crossed
polarizer is $I_{-} \sim \sin^2\left(\theta_{\rm rot}\right)$. We
therefore expect $I_{-}$ to jump discontinuously across any first
order transitions such as FM-AFM or XY-ASF phase boundaries and
gradually change across second order transitions such as the FM-a-SF
or XY-2SF phase boundaries. Of course, such measurements have to be
supplemented with momentum distribution function measurement to
distinguish between the superfluid and Mott phases.

A brief comment on system preparation and validity of use of grand
canonical ensemble is in order here. One should note that although
use of a grand canonical ensemble in the single species case is a
good approximation due to the presence of a confining potential 
(in this case 
the confining potential produces spatial inhomogeneities in the filling factor
and some of the outer regions will act like a particle reservoir for
the inner region),
this
is not necessarily the case in the two species system and other
means may have to be sought.

\section{Conclusions}
\label{conclusion}

In conclusion, we have studied the MI-SF transition in a
consisting of two species of ultracold atoms in an optical lattice in
a previously unexplored parameter region. We have used an ${\rm
O}(t/U)$ mean-field theory to explain the qualitative features of
the transition in most regions of the phase diagram.  This is
followed by incorporating the ${\rm O}(t^2/U^2)$ exchange effects
using a canonical transformation method and a quantum Monte Carlo
calculation.  All of these methods show that the
superfluid-insulator transition can occur with either depopulation
of species $2$ (a-SF phase) or simultaneous
onset of superfluidity of both species (SF-SF phase) or Mott
insulator of species $2$ coexisting with superfluid of species $1$
(${\rm MI}_2+{\rm SF}_1$ phase) and can be first order in
large regions of the phase diagram.
We have also shown that, whereas some qualitative features of the
SI transition can be obtained from ${\rm O}(t/U)$ mean-field
theory, incorporating the ${\rm O}(t^2/U^2)$ corrections is
necessary to deduce the
details
of the phase diagram and
order of the transitions between the phases. Our quantum Monte
Carlo study lends strong support to the above-mentioned results
and also shows screening of bosons of species $2$ in the SF-SF
phase. We also discussed possible experimental tests
of some of our predictions.
\begin{acknowledgments}

The authors thank A.M.S. Tremblay and O.F. Sylju{\aa}sen for helpful
discussions. KS and SMG were supported by NSA and ARDA under ARO
contract number DAAD19-02-1-0045 and the NSF through DMR-0342157. AI
was partly supported by The Swedish Foundation for International
Cooperation in Research and Higher Education (STINT). MCC was
supported by grant No. R05-2004-000-11004-0 from Korean Ministry of
Science and Technology. Monte Carlo calculations were in part
carried out using NorduGrid, a Nordic testbed for wide area
computing and data handling. KS also thanks Yong Baek Kim for
support through NSERC during completion of a part of this work and
Sergei Isakov for useful discussion.

\end{acknowledgments}

\section*{Appendix A}
Here we briefly sketch the derivation of $H^{\ast}$ (Eq.\
\ref{hamf}) starting from Eq.\ \ref{expansion}. To do this we show
the detailed derivation of two terms
$H_1 = [iS,H_0]$ and
$H_2=[iS,T_{\sigma}]$. The derivations of all other terms follow
in a similar fashion.

\subsection{$H_1$}
\label{h1}
To compute $H_1$, we use Eq.\ \ref{transop} to expand $S$ and write
\begin{eqnarray}
[iS,H_0]&=&-\frac{1}{\lambda U}\sum_\sigma\left[[\lambda
P_\sigma^S+P_\sigma^D, T_\sigma],H_0\right] \nonumber\\
&=& -\frac{1}{\lambda U}\sum_\sigma\Big [\left(\lambda P_\sigma^ST_\sigma
+P_\sigma^D T_\sigma \right.\nonumber\\
&& \left.-\lambda T_\sigma P_\sigma^S-T_\sigma P_\sigma^D\right),
\,H_0 \Big] \label{ex1}
\end{eqnarray}
Now consider the first term in the commutator in Eq.\ \ref{ex1}. Noting that
the projection operator $P_{\sigma}$ always
projects
onto the states $|1,0>$ or $|0,1>$, we see that we can write
\begin{eqnarray}
P_\sigma^ST_\sigma H_0 - H_0 P_\sigma^ST_\sigma
&=& U P_\sigma^ST_\sigma
\label{opid}
\end{eqnarray}
 This is an operator identity guaranteed by the construction of the projection
operator $P_{\sigma}$. Other terms in Eq.\ \ref{ex1} can be written in a
similar fashion and we have
\begin{eqnarray}
H_1 &=&-\sum_\sigma\left( P_\sigma^ST_\sigma
+P_\sigma^D T_\sigma + T_\sigma P_\sigma^S+T_\sigma
P_\sigma^D\right)\nonumber\\
&=&-\sum_\sigma\left( P_\sigma T_\sigma +
T_\sigma P_\sigma^S\right)=-\sum_\sigma T_\sigma^1
\label{firstcomm}
\end{eqnarray}
where in obtaining the last line we have again used the properties of the
projection operators $P_{\sigma}^S$. Combining Eqs.\ \ref{toperator},
\ref{firstcomm} and \ref{transop}, we get the second term in Eq.\
\ref{hamf}.

\subsection{$H_2$}
\label{h2}
Now we consider the term $H_2= [iS,\sum_{\sigma} T_{\sigma}^1]$. For this,
as we shall see, it is useful to define the operator ${\mathcal M}_{\sigma}=
P_{\sigma}T_{\sigma} + T_{\sigma} P_{\sigma}$. Then one can write, using
Eq.\ \ref{transop}
\begin{eqnarray}
H_2 &=&-\frac{1}{\lambda
U}\sum_{\sigma,\sigma^\prime}\left[ [P_\sigma^\lambda T_\sigma] ,
{\mathcal M}_{\sigma'}\right]
\label{h21}
\end{eqnarray}
Note that unlike $H_1$, here the sum extends over two different bonds
$\sigma$ and $\sigma'$. Consequently, expansion of Eq.\ \ref{h21} leads to
terms which can be classified into two categories.The first type of terms
involves two hopping operators $T_{\sigma}$ on the same bond while the
second involves the hopping operators on the different bonds:
\begin{eqnarray}
H_2 &=& H_{2a} + H_{2b} \nonumber\\
H_{2a} &=& -\frac{1}{\lambda U}\sum_{\sigma}\left[ [
P_\sigma^\lambda, T_\sigma] ,{\mathcal M}_{\sigma} \right]
\label{h2a1} \\
H_{2b} &=& \frac{1}{\lambda U}
\sum_{\sigma}\sum_{j}\left[ [P_\sigma^{\lambda}, T_\sigma] ,
{\mathcal M}_{\sigma+j}\right]
\label{h2b1}
\end{eqnarray}
where the sum over $j$ extend over the bonds which are nearest neighbors
to $\sigma$.

We first consider $H_{2a}$. We expand the operators ${\mathcal
M}_{\sigma}$ and $P_{\sigma}^{\lambda}$ and use the relation
$[P_{\sigma},P_{\sigma}^{\lambda}]=0$. Also, we note that all terms
of the form $P_{\sigma}^{1} T_{\sigma} P_{\sigma}^0 T_{\sigma}$ in
such an expansion vanish identically. After some straightforward
algebra, one obtains
\begin{eqnarray}
H_{2a} &=& -\frac{1}{\lambda U} \left[ P_{\sigma}^{\lambda} T_{\sigma}^2
P_{\sigma} +  P_{\sigma} T_{\sigma}^2
P_{\sigma}^{\lambda} -2 T_{\sigma} P_{\sigma}^{\lambda} T_{\sigma}
\right] \label{h2a2}
\end{eqnarray}
Note that the first two terms of Eq.\ \ref{h2a2}
represent
second order
virtual hopping processes and thus give the $t^2/U$ terms responsible for
the isospin ordering of the Mott phases, while the third term represents
two particle-hopping across a bond with an intermediate virtual state of
one particle on each side of the bond.

Next we come to computation of $H_{2b}$. We again expand out the
operators as before. Here, the crucial identity is that any terms of
the form $P_{\sigma} T_{\sigma} P_{\sigma + j}^{\lambda} T_{\sigma
+j}$ or $P_{\sigma+j} T_{\sigma+j} P_{\sigma}^{\lambda} T_{\sigma}$
vanish as long as $\sigma$ and $\sigma +j$ denotes nearest-neighbor
bonds. Using this, one gets
\begin{eqnarray}
H_{2b} &=& -\frac{1}{\lambda U}\sum_{\sigma}\sum_{j}
\Big ( P_\sigma^\lambda T_\sigma T_{\sigma+j}P_{\sigma+j} \nonumber\\
&& -T_\sigma P_\sigma^\lambda P_{\sigma+j} T_{\sigma+j}+{\rm h.c.}\Big)
\label{h2b2}
\end{eqnarray}

The other term $\left[iS,\left[iS,H_0\right]\right]$ in Eq.\ \ref{expansion}
can be computed in a similar fashion. Using all these results, we finally
obtain Eq.\ \ref{hamf}.



\begin{thebibliography}{99}
\bibitem{comment1} This way of formulating the mean-field theory can
be shown to be equivalent to the Hubbard-Stratonovitch decoupling
approach, provided the isospin flip process is taken into account in
computation of the on-site Green function.
\end{thebibliography}
\end{document}